\documentclass{article}

\usepackage{arxiv}

\usepackage[utf8]{inputenc} % allow utf-8 input
\usepackage[T1]{fontenc}    % use 8-bit T1 fonts
\usepackage{hyperref}

\usepackage{url}            % simple URL typesetting
\usepackage{booktabs}       % professional-quality tables
\usepackage{amsfonts}       % blackboard math symbols
\usepackage{nicefrac}       % compact symbols for 1/2, etc.
\usepackage{microtype}      % microtypography

\usepackage{graphicx}

\usepackage{booktabs}
\usepackage{array}
\usepackage{pdflscape}
\usepackage{multirow}

\usepackage{longtable}
\usepackage{rotating}
\usepackage{multicol}
\usepackage{graphicx}
\usepackage{float}

\title{Optimizing TESS-related Exoplanet Observation: A Systematic Approach to Scheduling JWST SOSS and BOTS Templates}

\author{\includegraphics[scale=0.06]{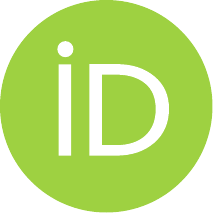}\hspace{1mm}Zoutong Shen}
\renewcommand{\shorttitle}{Optimizing TESS-related Exoplanet Observation}

\hypersetup{
pdftitle={Optimizing TESS-related Exoplanet Observation},
pdfsubject={astrophysics},
pdfauthor={Zoutong Shen},
pdfkeywords={JWST, TESS, Exoplanet},
}

\begin{document}
\maketitle

\begin{abstract}
This study presents a systematic approach to optimize the scheduling of exoplanet observations using the James Webb Space Telescope (JWST), focusing on targets discovered by the Transiting Exoplanet Survey Satellite (TESS). We developed a methodology to refine transit timing predictions for JWST's Cycle 3 Guest Observer program, specifically for the NIRISS/SOSS and NIRSpec/BOTS observation modes. Our process involved data collection from JWST proposal documents, cross-matching with TESS data, and applying the Transit Least Squares (TLS) algorithm for transit detection and characterization. We created comprehensive timelines for instrument usage and individual proposals, providing a visual representation of the observation schedule from July 2024 to September 2025. This approach demonstrates the potential for improved efficiency in JWST time allocation and sets a foundation for future refinements in astronomical observation planning.
\end{abstract}

% keywords can be removed

\section{Introduction}

The James Webb Space Telescope (JWST), launched on December 25, 2021, represents a monumental leap in our ability to study the universe, including exoplanets \cite{gardner_james_2006}. As a 6.5m diameter cold space telescope with cameras and spectrometers covering 0.6-28$\mu$m wavelengths, JWST extends the discoveries and technologies of its predecessors, the Hubble Space Telescope and the Spitzer Space Telescope \cite{mcelwain_james_2023}.

This project aims to develop an efficient scheduling system for JWST's Guaranteed Time Observations (GO3) program, specifically targeting exoplanets discovered by the Transiting Exoplanet Survey Satellite (TESS) \cite{ricker_transiting_2015}. Our focus is on optimizing the use of two key JWST observation modes: Bright Object Time Series (BOTS) and Single Object Slitless Spectroscopy (SOSS), which are particularly suited for exoplanet characterization.

JWST's unique capabilities make it an ideal tool for exoplanet research:

\begin{itemize}
    \item Its infrared wavelength range penetrates dust clouds, allowing observation of obscured planetary systems and star formation regions \cite{Rigby_2023}.
    \item It can detect objects too cool to radiate visible light, including the fundamental vibration-rotation bands of important molecules in exoplanet atmospheres.
    \item JWST's collecting area is significantly larger than previous space telescopes, with Hubble having only 1/6.25 of JWST's collecting area, allowing for unprecedented sensitivity \cite{Rigby_2023}.
    \item Its location at the Sun-Earth L2 point and advanced cooling systems enable highly stable observations in the infrared spectrum \cite{claimpin_overview_2010.7731E..07C}.
\end{itemize}

The BOTS mode, available on both NIRCam and NIRSpec, is designed for high-precision time-series observations of bright sources. NIRCam BOTS uses rapid readout of subarrays to increase the readout cadence and increase the saturation limits \cite{rieke_performance_2023}. This mode is crucial for observing planetary transits and secondary eclipses, providing detailed information about exoplanet atmospheres.

NIRISS SOSS mode provides medium-resolution (R$\sim$700) spectroscopy optimized for time-series observations of transiting exoplanets \cite{doyon_near_2023}. It uses a unique crossed-dispersed grism that spreads the spectrum over many detector pixels, allowing for high photometric precision and wavelength coverage from 0.6 to 2.8$\mu$m in a single observation.

Our scheduling system will leverage these advanced capabilities to maximize the scientific output of exoplanet observations. We will consider factors such as target visibility, instrument modes, and observation duration to create an optimal observation plan for TESS-discovered exoplanets.

This work is particularly significant given JWST's projected fuel-limited lifetime of up to 20 years and its unique observational capabilities that cannot be obtained in any other way. By efficiently scheduling TESS follow-up observations using BOTS and SOSS modes, we aim to contribute significantly to our understanding of exoplanet diversity, atmospheric compositions, and potential habitability, advancing our quest to contextualize Earth's place in the universe and potentially detect Earth-like planets around Sun-like stars.
\section{Methodology}
\subsection{Data Collection and Initial Processing} 

This study focuses on TESS follow-up observations, specifically targeting transiting exoplanets. To gather relevant data, we accessed the James Webb Space Telescope (JWST) Cycle 3 Guest Observer (GO3) program website, concentrating on the ``Exoplanets and Exoplanet Formation'' category. Our analysis primarily centers on two instrument modes: NIRISS/SOSS (Near Infrared Imager and Slitless Spectrograph/Single Object Slitless Spectroscopy) and NIRSpec/BOTS (Near Infrared Spectrograph/Bright Object Time Series). These modes are particularly well-suited for exoplanet transit observations due to their high precision and ability to capture time-series spectroscopic data \cite{sullivan_trans_2015, stevenson_2016}.

Our initial query yielded 19 proposals encompassing 66 unique missions that met our criteria. Within this dataset, 40 observations utilize NIRSpec/BOTS, while 24 employ NIRISS/SOSS. It's worth noting that some missions investigate the same celestial object using multiple instruments beyond the scope of this study. Consequently, we filtered out related observations using MIRI (Mid-Infrared Instrument) and NIRCam (Near Infrared Camera), as they fall outside our primary focus on transit spectroscopy.

To refine and validate future transits, we cross-matched our JWST targets with the TESS database using Lightkurve objects \cite{Lightkurve2018}. By setting the cadence parameter to 120 seconds, we ensured the acquisition of high-quality light curves, which are crucial for precise transit timing measurements \cite{hord_uniform_2021}. This process resulted in a final dataset of 39 unique observation targets, each associated with their respective TICID (TESS Input Catalog Identifier), TESS magnitude, Right Ascension (RA), and Declination (Dec). These parameters are essential for future timing computations, particularly when working with Heliocentric Julian Dates (HJD).

\begin{figure}[ht]
    \centering
    \includegraphics[width=0.8\textwidth]{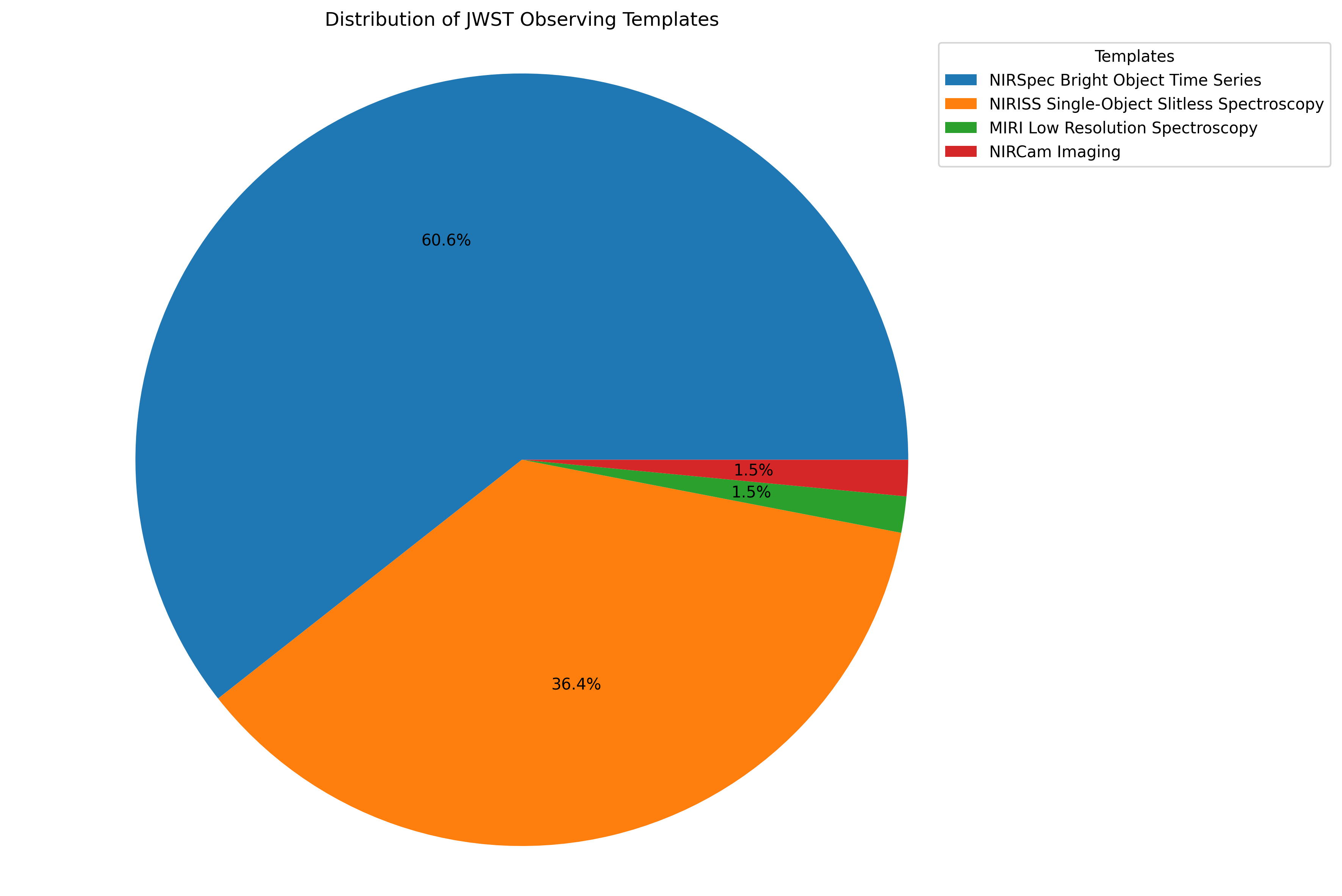}
    \caption{Distribution of JWST observing templates in the final dataset.}
    \label{fig:template_distribution}
\end{figure}

Figure \ref{fig:template_distribution} presents a pie chart illustrating the distribution of observing templates in our final dataset. This visualization underscores the prevalence of NIRSpec/BOTS and NIRISS/SOSS modes in our study, reflecting their importance in exoplanet transit observations.

The list of cross-matched JWST objects, including their TICID, TESS magnitude, RA, and Dec, is documented. As shown in Table~\ref{tab:jwst_targets} in the Appendix, we have a total of 34 JWST targets with corresponding TESS data. This comprehensive dataset forms the foundation for our subsequent analysis of exoplanet transit timing and characterization.

\subsection{Data Filtering and Refinement}

Following the initial data collection, we implemented a rigorous filtering process to refine our observational dataset. This process ensured that our final target list comprised only the most suitable candidates for our analysis. Our refinement procedure involved several key steps to enhance the quality and reliability of our dataset.

First, we assessed the periodicity of each target in our initial list. Utilizing the Transit Least Squares (TLS) algorithm \cite{hippke_tls_2019}, we evaluated the periodic nature of the light curves. Targets without discernible periodicity or those with extremely long periods that exceeded the operational constraints of the TLS algorithm were excluded from further analysis. This step was crucial in ensuring that our dataset contained only targets with reliable, detectable transit signals.

Next, we conducted a thorough review of each mission's status by accessing the official JWST Guaranteed Time Observations (GTO) and General Observers (GO) program websites. For each target, we retrieved and examined the public PDF documents containing detailed mission descriptions. We paid particular attention to the implementation section of each proposal, which typically includes the mission status, planned observation date, and planned observation duration. Proposals with a status of "archived" or "withdrawn" were removed from our dataset, focusing our analysis on missions either implemented or scheduled for future observation.

The final step in our refinement process involved consolidating the information gathered from the previous steps. We created a comprehensive list that included the target's zero-phase time in Heliocentric Julian Date (HJD) as specified in the mission descriptions, the predicted orbital period of each exoplanet, and the planned observation date and duration for each target. This refined list excluded any missions lacking specific assigned observation dates.

Through this meticulous filtering process, we significantly reduced our initial dataset to a refined list of high-quality targets. We excluded observations with clearly defined observation period, since that means our prediction will not contribute to the scheduling, removed archived or withdrawn proposals, and ensured that each remaining target had well-defined observation parameters. This refined list forms the foundation for our subsequent analysis, ensuring that we focus on the most promising and well-documented exoplanet candidates for this work's JWST scheduling mission.

\subsection{Transit Prediction and Validation}

The final stage of our methodology involves the prediction and validation of transit events for our refined target list. This process is crucial for accurately timing JWST observations and maximizing the scientific output of each observation window.

\subsubsection{Light Curve Retrieval and Processing}

We utilized the Lightkurve package \cite{Lightkurve2018} to retrieve light curves from TESS for each target in our refined list. Lightkurve offers a Python-based toolkit for Kepler and TESS time series analysis, providing efficient access to and manipulation of light curve data. 

Prior to analysis, we performed data cleaning and normalization. This process included removing outliers, detrending the light curve to account for long-term variations, and normalizing the flux values. Importantly, we removed gaps between TESS observation sectors to ensure continuous data for accurate transit duration estimation.

\subsubsection{Transit Detection with TLS}

For transit detection and characterization, we employed the Transit Least Squares (TLS) algorithm \cite{hippke_tls_2019}. TLS offers several advantages over traditional box-fitting algorithms, including improved sensitivity to shallow transits and more robust handling of stellar limb darkening. However, it can be computationally intensive for very long period systems.

The TLS algorithm was applied to each processed light curve, yielding key transit parameters:
\begin{itemize}
    \item Transit epoch (T$_0$)
    \item Orbital period
    \item Transit depth
    \item Transit duration
    \item Signal Detection Efficiency (SDE)
\end{itemize}

\begin{figure}[htbp]
    \centering
    \includegraphics[width=\textwidth]{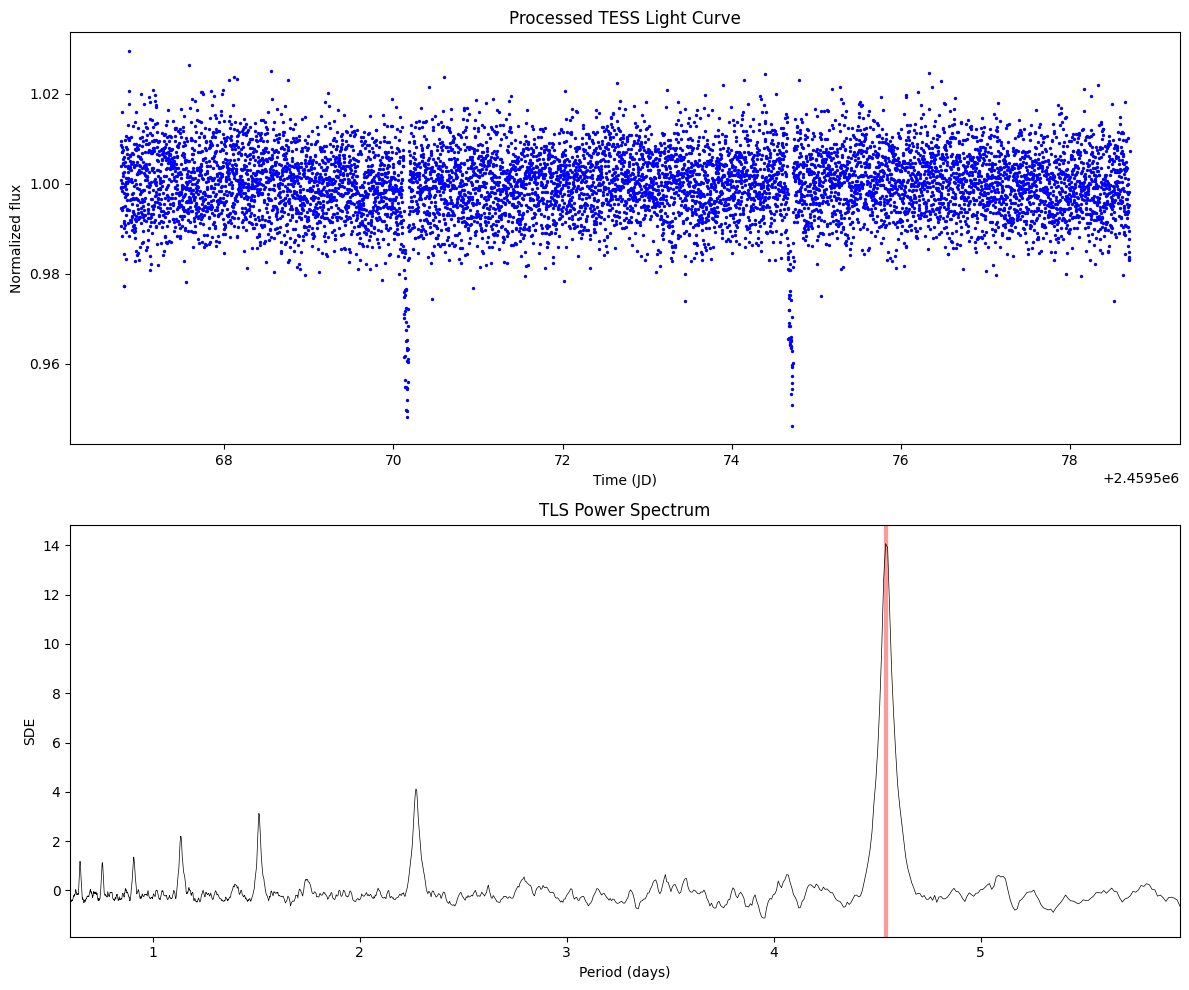}
    \caption{Example of TLS analysis for proposal ID 583, observation 1. Top: Processed TESS light curve. Bottom: TLS power spectrum showing the detected period.}
    \label{fig:tls_example}
\end{figure}

Figure \ref{fig:tls_example} illustrates the TLS output for proposal ID 583, observation number 1, showcasing both the processed TESS light curve and the resulting TLS power spectrum.

\subsubsection{Validation and Refinement}

To validate our TLS results, we compared the derived periods with those provided in the JWST publicly available proposal documents. For cases where the TLS-derived period deviated by more than 10\% from the proposal value, we adopted a more cautious approach. In these instances, we utilized the start and end phase information from the proposal PDFs, along with the zero-phase reference time (converted from HJD to BJD using the target's RA and DEC) to recalculate future transit events.

This discrepancy could arise from several factors:
\begin{itemize}
    \item Harmonics in the system interfering with the TLS algorithm
    \item Extremely long orbital periods beyond the sensitivity of TLS given the available TESS data
    \item Complex multi-planet systems where signals may overlap
\end{itemize}

\subsubsection{Transit Prediction}

Using the validated orbital parameters, we implemented a precise transit prediction process:

\begin{enumerate}
    \item We calculated the three closest transits to the NASA-assigned date for each target as listed on the GO3 website. For each predicted transit, we documented:
    \begin{itemize}
        \item Transit start time
        \item Transit end time
        \item Transit midpoint (T$_0$)
    \end{itemize}
    
    \item For each predicted transit, we created an observation window centered on T$_0$. The duration of this window was based on the transit duration determined by TLS or provided in the proposal.
    
    \item We then selected the transit whose observation window was closest to the JWST-assigned date. For this comparison, we assumed a timestamp of 00:00 for the NASA-assigned date if no specific time was provided.
\end{enumerate}

This approach ensures that our predicted observation window:
\begin{itemize}
    \item Aligns closely with NASA's scheduling plans
    \item Captures the full transit event
    \item Maximizes the scientific potential of the JWST observation time
\end{itemize}

By predicting multiple transits and selecting the most appropriate one, we provide flexibility in scheduling while maintaining optimal transit coverage. This method accounts for potential conflicts or constraints in JWST's observing schedule, allowing for alternative observation windows if needed.

\section{Results}

Our analysis of the transit observation schedule for the entire GO3 is represented by two key sets of visualizations: instrument timelines and individual proposal timelines.

\subsection{Instrument Timelines}

The first set of images displays the timelines for two primary instruments: BOTS (Bright Object Time Series) and SOSS (Single Object Slitless Spectroscopy). These timelines span the entire GO3 period and provide a high-level overview of the observation schedule.

\begin{sidewaysfigure}
    \centering
    \includegraphics[width=\textheight,height=\textwidth,keepaspectratio]{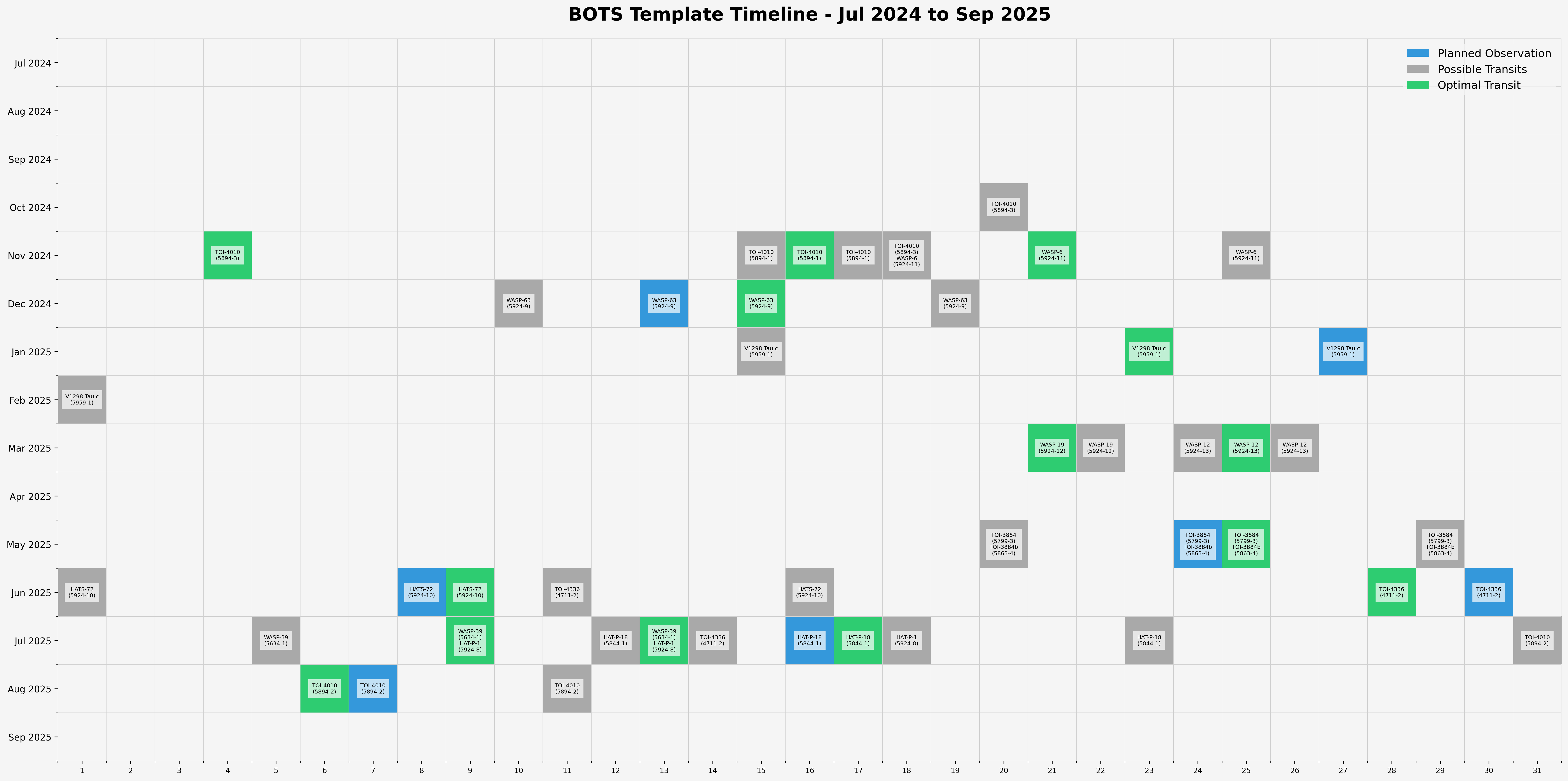}
    \caption{BOTS Template Timeline - GO3}
    \label{fig:bots_timeline}
\end{sidewaysfigure}

\begin{sidewaysfigure}
    \centering
    \includegraphics[width=\textheight,height=\textwidth,keepaspectratio]{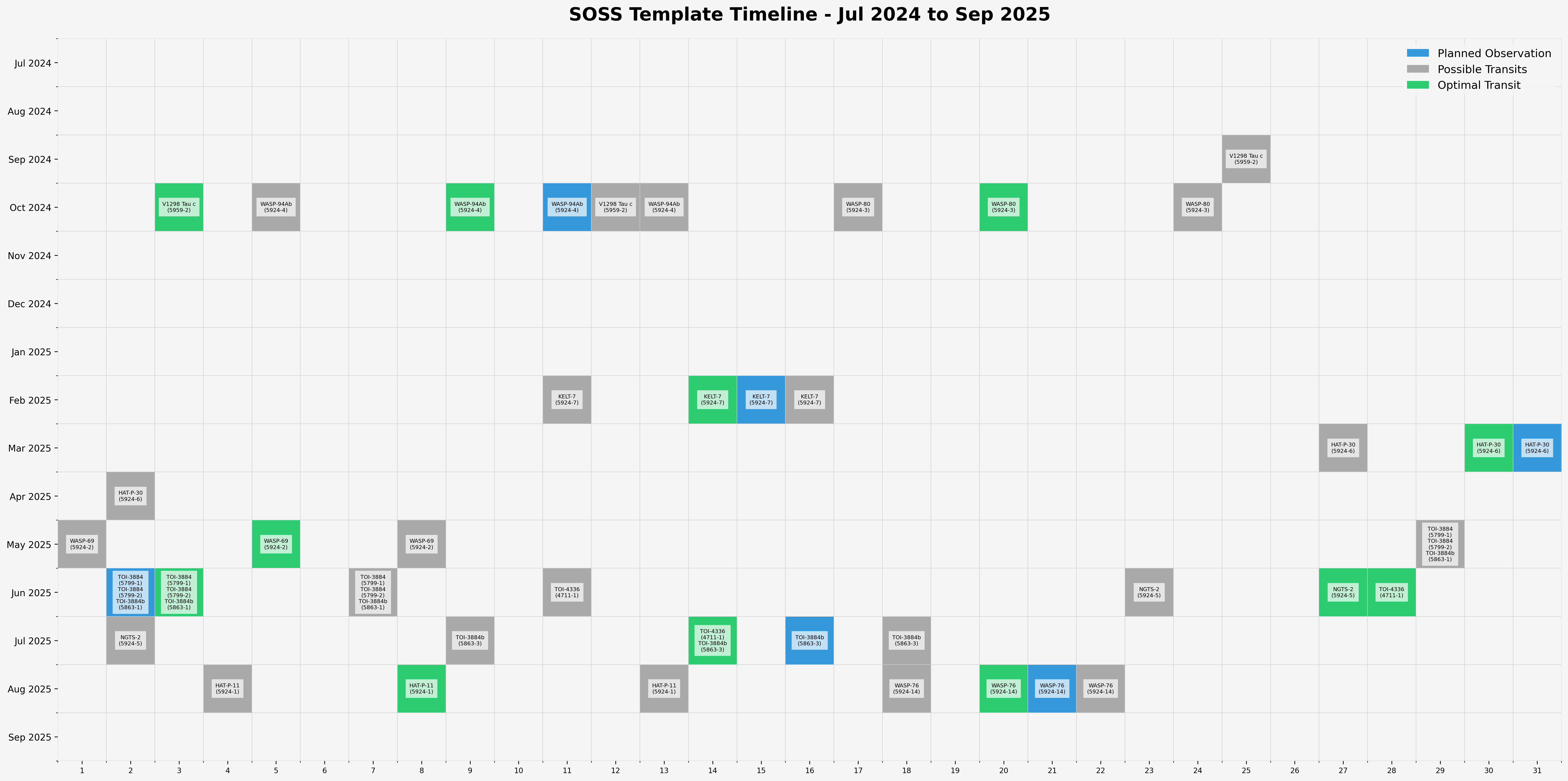}
    \caption{SOSS Template Timeline - GO3}
    \label{fig:soss_timeline}
\end{sidewaysfigure}

Key features of these timelines:
\begin{itemize}
    \item Color-coded representation: Blue for planned observations, grey for possible transits, and green for optimal transits
    \item Monthly breakdown allowing for easy identification of busy periods and gaps in the schedule
    \item Visual distribution of activities across the entire observation period for each instrument
\end{itemize}

These timelines enable quick identification of periods with high observational activity, potential conflicts, and optimal observation windows for each instrument. They also highlight any significant gaps in the schedule, allowing for efficient resource allocation and planning.

\subsection{Individual Proposal Timelines}

The second set of images represents the specific timing information for individual proposals. These timelines provide a more granular view of each observation within a proposal. For instance, one such timeline might show multiple observations for a single target, each with its own optimal window and possible transit periods.

\begin{figure}[htbp]
    \centering
    \includegraphics[width=\textwidth]{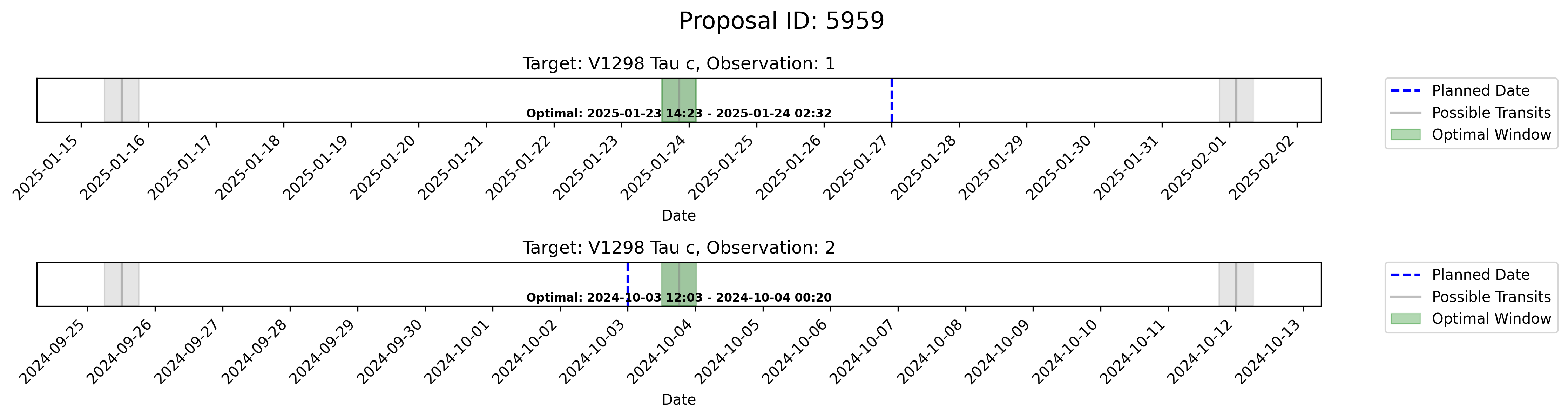}
    \caption{Sample Individual Proposal Timeline - Proposal ID: 5959, Target: V1298 Tau c}
    \label{fig:sample_proposal_timeline}
\end{figure}

Key elements of these timelines:
\begin{itemize}
    \item Separate panels for each observation within a proposal
    \item Clear indication of the planned observation date (blue dashed line)
    \item Visualization of possible transit windows (grey bars)
    \item Highlighted optimal observation window (green bar)
    \item Precise timing information for each element
\end{itemize}

These detailed timelines allow for precise planning of individual observations, taking into account the optimal observation windows and potential transit times. They provide crucial information for scheduling and maximizing the scientific output of each proposal.

Together, these two sets of visualizations offer a comprehensive view of the observation schedule, from the broad instrument-level perspective down to the specific timing of individual proposals. This multi-level representation allows for both strategic planning and detailed execution of the transit observation program.

\textit{Note: Detailed timelines for all individual proposals are included in the Appendix for reference.}

\section{Discussion}

The James Webb Space Telescope (JWST) represents a significant investment in observational astronomy, with each cycle offering precious observation time. Our primary objective is to optimize the scheduling process, thereby maximizing the telescope's utility for the scientific community. This study represents an initial step towards this goal, focusing on refining transit timing predictions using TESS data.

While TESS was initially conceived as a tool for target selection for JWST, our work demonstrates its additional value in refining orbital parameters crucial for scheduling. These include period calculation, transit duration estimation, and future event prediction. These refined parameters enable more precise scheduling of JWST observations, potentially increasing the efficiency of telescope time allocation.

The methodology developed in this study has potential for broader application. It could be adapted to optimize scheduling for other ground-based and space-based telescopes, creating a more comprehensive framework for astronomical observation planning. For SOSS and BOTS instruments specifically, future studies could extract and analyze Full Frame Images using tools like \texttt{eleanor} (\cite{feinstein_eleanor_2019})with a 30-minute cadence rate, potentially providing even more detailed information for scheduling optimization.

To further increase accuracy and retrieve additional planetary parameters, more sophisticated modeling approaches could be employed. Utilizing packages such as \texttt{batman} or \texttt{exoplanet} for detailed transit fitting could incorporate more complex physical models to extract additional planetary parameters. However, it's important to note that such advanced modeling is computationally expensive. For the current GO3 cycle, the urgency of schedule publication outweighed the potential benefits of these more time-consuming analyses.

For future observation cycles (GO4 and beyond), we recommend allocating more time for pre-observation preparation and conducting more detailed and comprehensive analyses. This could potentially include incorporating advanced modeling techniques as computational resources and time allow. These steps could further refine our scheduling optimization, potentially leading to even more efficient use of JWST's observational capabilities.

In conclusion, while our current approach provides a significant improvement in scheduling optimization, there remains considerable potential for further refinement and expansion of these techniques in future work. By continuing to leverage tools like TESS and exploring more advanced modeling techniques, we can work towards fully utilizing the power of JWST and similar instruments, maximizing their scientific output for the astronomical community.

Copy

\clearpage
\appendix
\section{List of JWST Targets}

\begin{longtable}{ccccc}
\caption{JWST Targets with TESS Data}\label{tab:jwst_targets} \\
\hline
JWST Target & TIC ID & TESS Magnitude & RA & Dec \\
\hline
\endfirsthead
\multicolumn{5}{c}{Table \ref{tab:jwst_targets} continued} \\
\hline
JWST Target & TIC ID & TESS Magnitude & RA & Dec \\
\hline
\endhead
\hline
\multicolumn{5}{r}{Continued on next page} \\
\endfoot
\hline
\endlastfoot
TOI-4336 & 166184428 & 11.0196 & 206.106157075407 & -40.3376476972936 \\
TOI-2525 b & 149601126 & 13.4006 & 86.8507833506736 & -60.5214083883845 \\
TOI-2525 c & 149601126 & 13.4006 & 86.8507833506736 & -60.5214083883845 \\
WASP-103 & 276754403 & 11.7586 & 249.314898846362 & 7.18336386302314 \\
Kepler-167 & 137686948 & 13.3685 & 292.658442551718 & 38.3453990625523 \\
WASP-39 & 181949561 & 11.3763 & 217.326729902534 & -3.44450084614399 \\
TOI-3884 & 86263325 & 12.9113 & 181.572654395924 & 12.5069127343743 \\
HAT-P-18 & 21744120 & 11.733 & 256.346448270061 & 33.0124830959209 \\
TOI-3884b & 86263325 & 12.9113 & 181.572654395924 & 12.5069127343743 \\
TOI-4010 & 352682207 & 11.5163 & 20.2144096197728 & 66.0722886403418 \\
HAT-P-11 & 28230919 & 8.5077 & 297.708364402628 & 48.0808603792123 \\
WASP-80 & 243921117 & 10.3622 & 303.167371561316 & -2.14421976578892 \\
WASP-94Ab & 92352620 & 9.6433 & 313.783102119552 & -34.1355576898925 \\
NGTS-2 & 125739286 & 10.5288 & 215.122871141882 & -31.2020661002262 \\
HAT-P-30 & 455135327 & 9.93908 & 123.949918879321 & 5.83676446194184 \\
KELT-7 & 367366318 & 8.14847 & 78.2955364801986 & 33.3181675626516 \\
HAT-P-1 & 346338552 & 9.76031 & 344.445184588974 & 38.6750987181784 \\
WASP-63 & 393414358 & 10.439 & 94.3364525276857 & -38.3232650531597 \\
HATS-72 & 188570092 & 11.3875 & 339.026329121847 & -16.9999411754881 \\
WASP-6 & 204376737 & 11.4779 & 348.157236940747 & -22.6739659690223 \\
WASP-19 & 35516889 & 11.6044 & 148.416985368292 & -45.6591822165323 \\
WASP-12 & 86396382 & 11.0967 & 97.6366528287122 & 29.6722962039341 \\
WASP-76 & 293435336 & 9.02706 & 26.6327402944207 & 2.70056477155313 \\
WASP-69 & 248853232 & 8.8628 & 315.025820520989 & -5.09445472449585 \\
V1298 Tau c & 15756231 & 9.4901 & 61.3316300572137 & 20.1571009815486 \\
V1298 Tau d & 15756231 & 9.4901 & 61.3316300572137 & 20.1571009815486 \\
TOI 451 c & 257605131 & 10.2666 & 62.9664459171879 & -37.9397832374178 \\
TOI 451 d & 257605131 & 10.2666 & 62.9664459171879 & -37.9397832374178 \\
TOI 2076 b & 27491137 & 8.3745 & 217.392678387443 & 39.7904291644655 \\
TOI 2076 c & 27491137 & 8.3745 & 217.392678387443 & 39.7904291644655 \\
TOI 2076 d & 27491137 & 8.3745 & 217.392678387443 & 39.7904291644655 \\
TOI-849 & 33595516 & 11.5485 & 28.7154332127702 & -29.4217975356229 \\
TRAPPIST-1 & 278892590 & 13.8529 & 346.622497498953 & -5.04134345727205 \\
WASP-76b & 293435336 & 9.02706 & 26.6327402944207 & 2.70056477155313 \\
\end{longtable}
\clearpage

\section{Observation Timing}

\begin{table}[htbp]
\centering
\caption{SOSS Observations}
\begin{tabular}{ccccccc}
\toprule
Proposal-Obs & Target & Planned Date & Duration & Closest Transit & \multicolumn{2}{c}{Optimal Window} \\
\cmidrule(lr){6-7}
& & & (hours) & & Start & End \\
\midrule
4711-1 & TOI-4336 & 2025-06-28 & 8.50 & 2025-06-28 07:46:31 & \textbf{2025-06-28 03:31:31} & \textbf{2025-06-28 12:01:31} \\
5799-1,2 & TOI-3884 & 2025-06-02 & 7.31 & 2025-06-03 03:58:25 & \textbf{2025-06-03 00:19:07} & \textbf{2025-06-03 07:37:43} \\
5863-1 & TOI-3884b & 2025-06-02 & 7.31 & 2025-06-03 03:58:25 & \textbf{2025-06-03 00:19:07} & \textbf{2025-06-03 07:37:43} \\
5863-3 & TOI-3884b & 2025-07-16 & 7.31 & 2025-07-14 02:07:22 & \textbf{2025-07-13 22:28:04} & \textbf{2025-07-14 05:46:40} \\
5924-1 & HAT-P-11 & 2025-08-08 & 7.78 & 2025-08-08 23:51:04 & \textbf{2025-08-08 19:57:40} & \textbf{2025-08-09 03:44:28} \\
5924-2 & WASP-69 & 2025-05-05 & 7.03 & 2025-05-05 00:16:47 & \textbf{2025-05-04 20:45:53} & \textbf{2025-05-05 03:47:41} \\
5924-3 & WASP-80 & 2024-10-20 & 7.25 & 2024-10-20 23:28:23 & \textbf{2024-10-20 19:50:53} & \textbf{2024-10-21 03:05:53} \\
5924-4 & WASP-94Ab & 2024-10-11 & 11.51 & 2024-10-09 02:33:09 & \textbf{2024-10-08 20:47:51} & \textbf{2024-10-09 08:18:27} \\
5924-5 & NGTS-2 & 2025-06-27 & 11.86 & 2025-06-27 18:15:42 & \textbf{2025-06-27 12:19:54} & \textbf{2025-06-28 00:11:30} \\
5924-6 & HAT-P-30 & 2025-03-31 & 7.47 & 2025-03-30 11:31:40 & \textbf{2025-03-30 07:47:34} & \textbf{2025-03-30 15:15:46} \\
5924-7 & KELT-7 & 2025-02-15 & 9.84 & 2025-02-14 06:09:14 & \textbf{2025-02-14 01:14:02} & \textbf{2025-02-14 11:04:26} \\
5924-14 & WASP-76 & 2025-08-21 & 9.89 & 2025-08-20 12:19:19 & \textbf{2025-08-20 07:22:37} & \textbf{2025-08-20 17:16:01} \\
\bottomrule
\end{tabular}
\end{table}

\begin{table}[htbp]
\centering
\caption{BOTS Observations}
\begin{tabular}{ccccccc}
\toprule
Proposal-Obs & Target & Planned Date & Duration & Closest Transit & \multicolumn{2}{c}{Optimal Window} \\
\cmidrule(lr){6-7}
& & & (hours) & & Start & End \\
\midrule
4711-2 & TOI-4336 & 2025-06-30 & 8.41 & 2025-06-28 07:46:31 & \textbf{2025-06-28 03:34:13} & \textbf{2025-06-28 11:58:49} \\
5634-1 & WASP-39 & 2025-07-09 & 7.98 & 2025-07-09 18:13:36 & \textbf{2025-07-09 14:14:12} & \textbf{2025-07-09 22:13:00} \\
5799-3 & TOI-3884 & 2025-05-24 & 7.23 & 2025-05-25 01:43:06 & \textbf{2025-05-24 22:06:12} & \textbf{2025-05-25 05:20:00} \\
5844-1 & HAT-P-18 & 2025-07-16 & 7.28 & 2025-07-17 19:27:32 & \textbf{2025-07-17 15:49:08} & \textbf{2025-07-17 23:05:56} \\
5863-4 & TOI-3884b & 2025-05-24 & 7.23 & 2025-05-25 01:43:06 & \textbf{2025-05-24 22:06:12} & \textbf{2025-05-25 05:20:00} \\
5894-1 & TOI-4010 & 2024-11-16 & 6.22 & 2024-11-16 10:38:37 & \textbf{2024-11-16 07:32:01} & \textbf{2024-11-16 13:45:13} \\
5894-2 & TOI-4010 & 2025-08-07 & 8.32 & 2025-08-06 00:32:40 & \textbf{2025-08-05 20:23:04} & \textbf{2025-08-06 04:42:16} \\
5894-3 & TOI-4010 & 2024-11-04 & 10.06 & 2024-11-04 04:01:42 & \textbf{2024-11-03 22:59:54} & \textbf{2024-11-04 09:03:30} \\
5924-8 & HAT-P-1 & 2025-07-13 & 8.76 & 2025-07-13 23:17:10 & \textbf{2025-07-13 18:54:22} & \textbf{2025-07-14 03:39:58} \\
5924-9 & WASP-63 & 2024-12-13 & 14.34 & 2024-12-15 01:48:50 & \textbf{2024-12-14 18:38:38} & \textbf{2024-12-15 08:59:02} \\
5924-10 & HATS-72 & 2025-06-08 & 8.02 & 2025-06-09 03:05:38 & \textbf{2025-06-08 23:05:02} & \textbf{2025-06-09 07:06:14} \\
5924-11 & WASP-6 & 2024-11-21 & 8.19 & 2024-11-21 19:37:23 & \textbf{2024-11-21 15:31:41} & \textbf{2024-11-21 23:43:05} \\
5924-12 & WASP-19 & 2025-03-22 & 5.93 & 2025-03-21 22:34:35 & \textbf{2025-03-21 19:36:41} & \textbf{2025-03-22 01:32:29} \\
5924-13 & WASP-12 & 2025-03-26 & 8.77 & 2025-03-25 21:11:44 & \textbf{2025-03-25 16:48:38} & \textbf{2025-03-26 01:34:50} \\
\bottomrule
\end{tabular}
\end{table}

\clearpage
\section{Appendix: All Proposal Timelines}

\begin{figure}[htbp]
    \centering
    \includegraphics[width=0.8\textwidth]{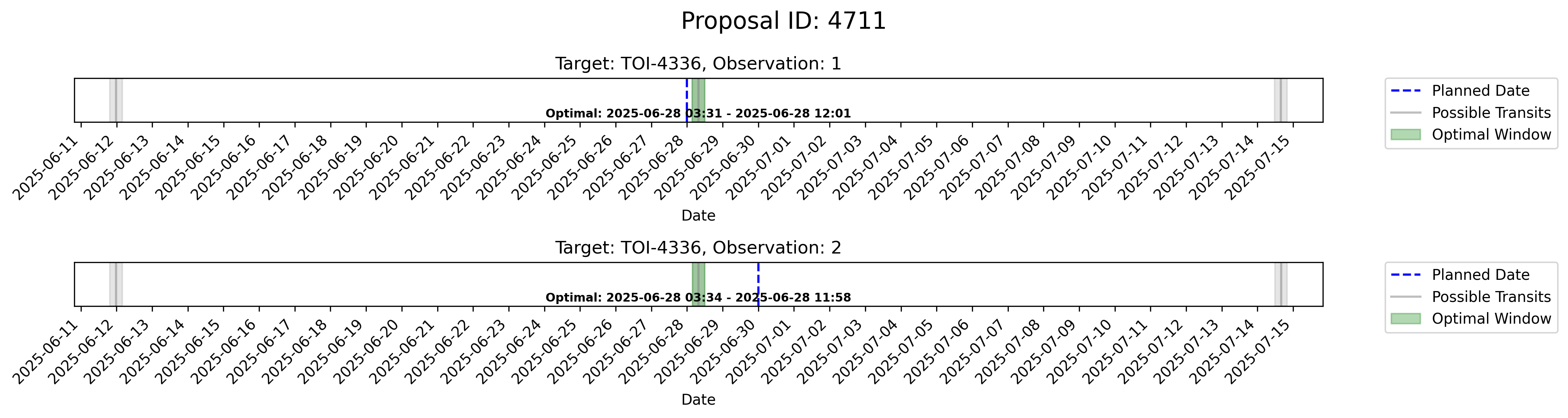}
    \caption{Timeline for Proposal 4711}
\end{figure}

\begin{figure}[htbp]
    \centering
    \includegraphics[width=0.8\textwidth]{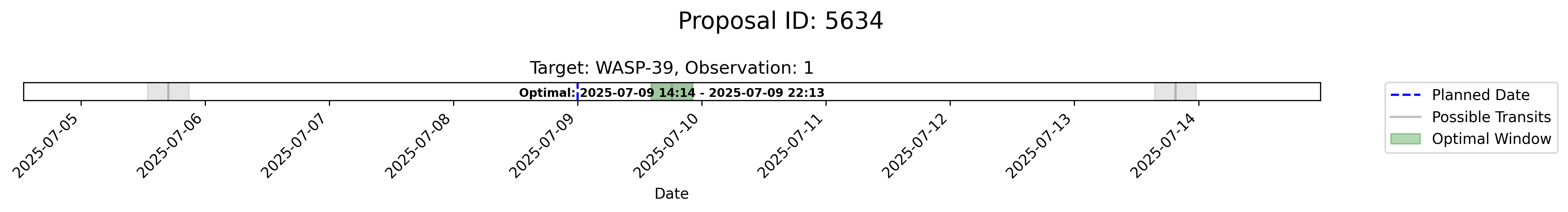}
    \caption{Timeline for Proposal 5634}
\end{figure}

\begin{figure}[htbp]
    \centering
    \includegraphics[width=0.8\textwidth]{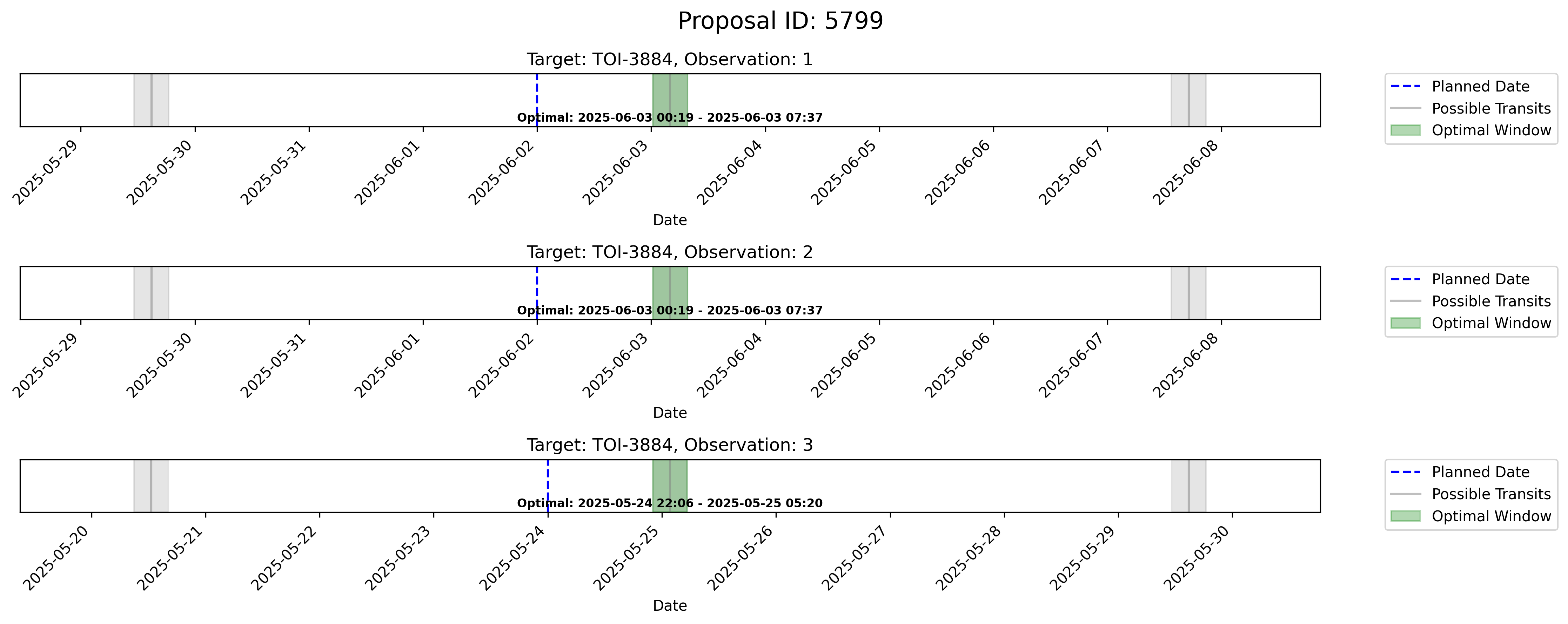}
    \caption{Timeline for Proposal 5799}
\end{figure}

\begin{figure}[htbp]
    \centering
    \includegraphics[width=0.8\textwidth]{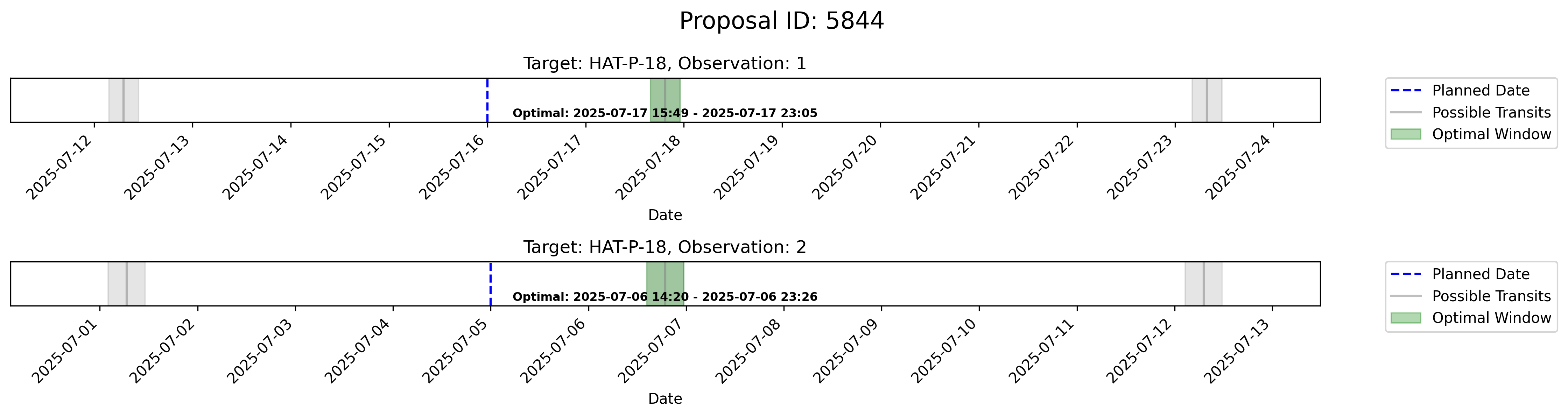}
    \caption{Timeline for Proposal 5844}
\end{figure}

\begin{figure}[htbp]
    \centering
    \includegraphics[width=0.8\textwidth]{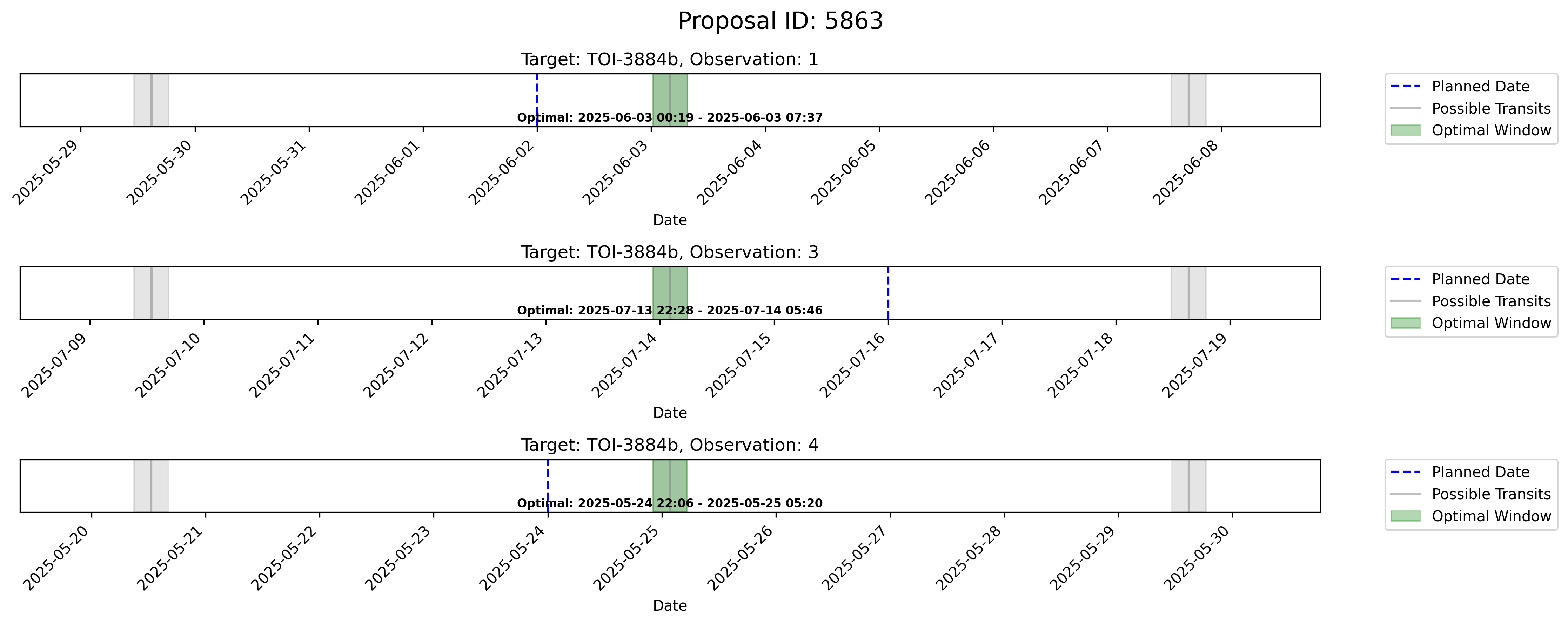}
    \caption{Timeline for Proposal 5863}
\end{figure}

\begin{figure}[htbp]
    \centering
    \includegraphics[width=0.8\textwidth]{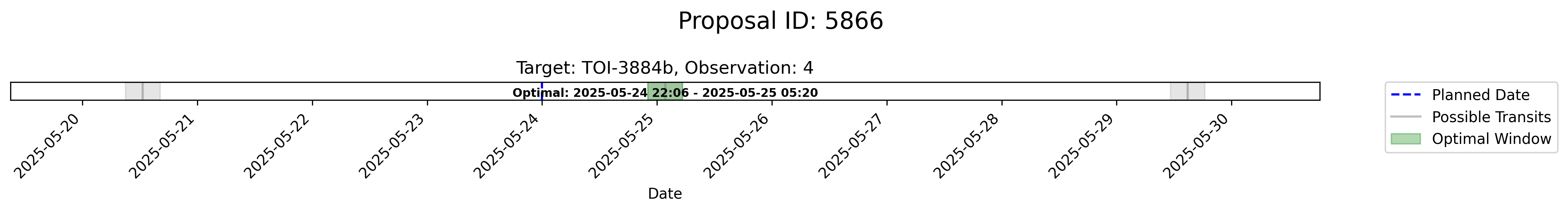}
    \caption{Timeline for Proposal 5866}
\end{figure}

\begin{figure}[htbp]
    \centering
    \includegraphics[width=0.8\textwidth]{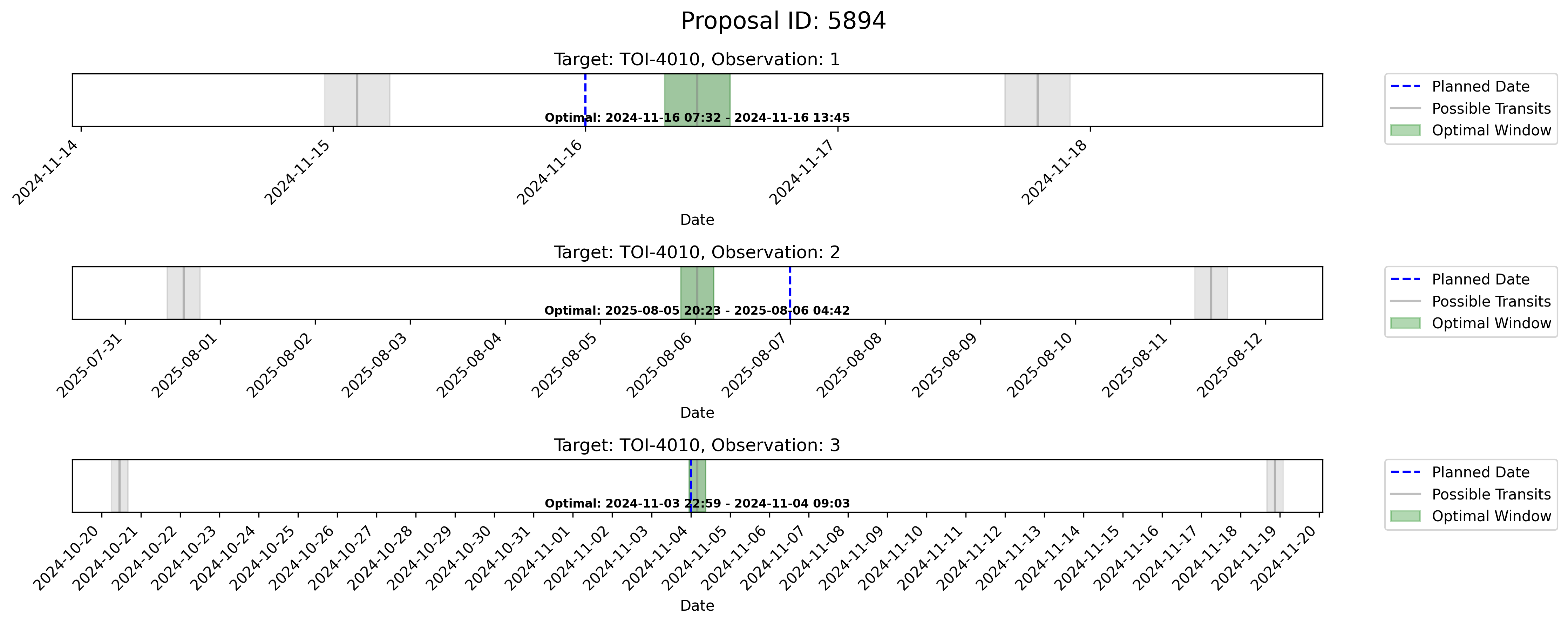}
    \caption{Timeline for Proposal 5894}
\end{figure}

\begin{figure}[htbp]
    \centering
    \includegraphics[width=0.8\textwidth]{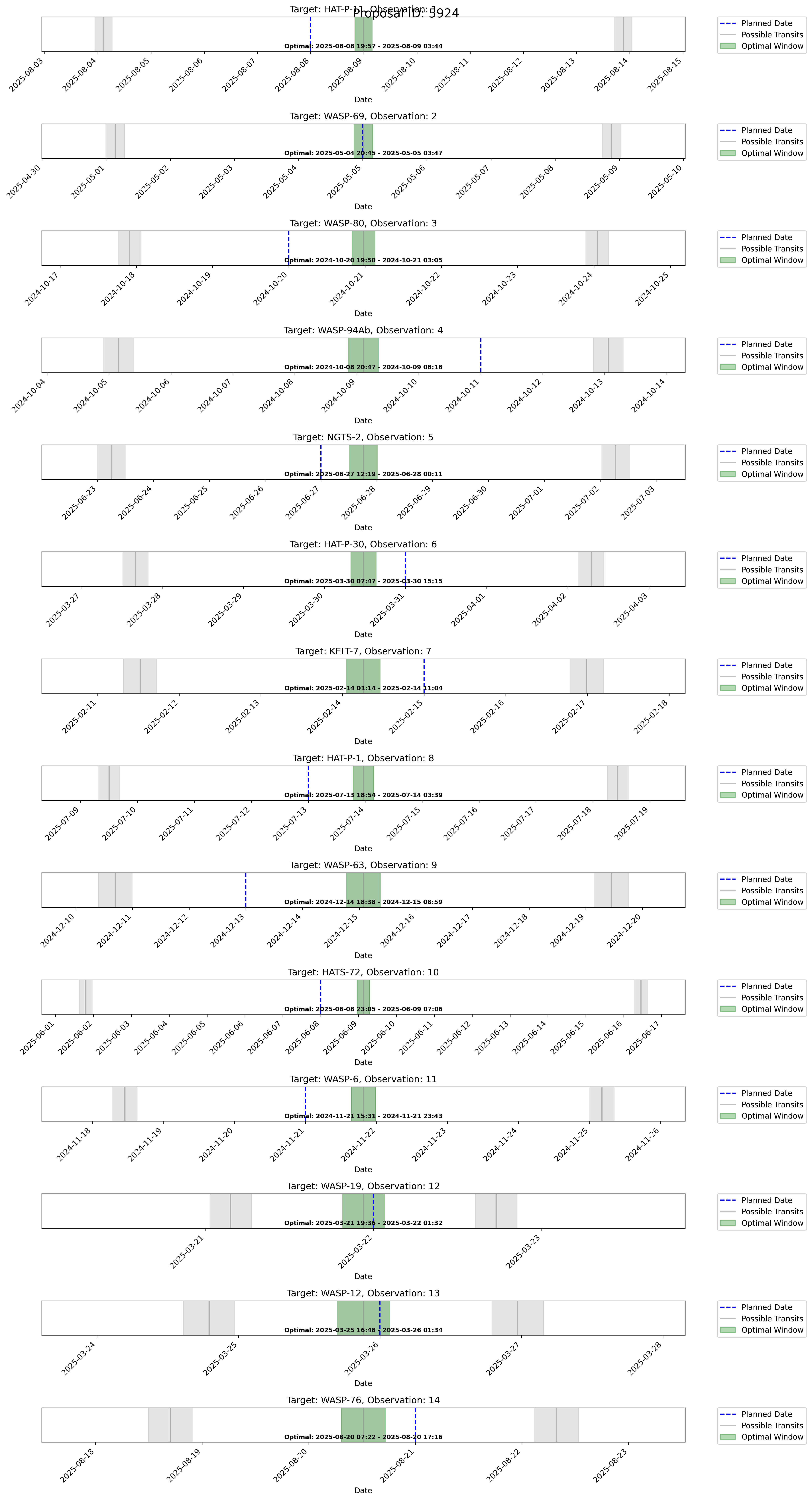}
    \caption{Timeline for Proposal 5924}
\end{figure}

\begin{figure}[htbp]
    \centering
    \includegraphics[width=0.8\textwidth]{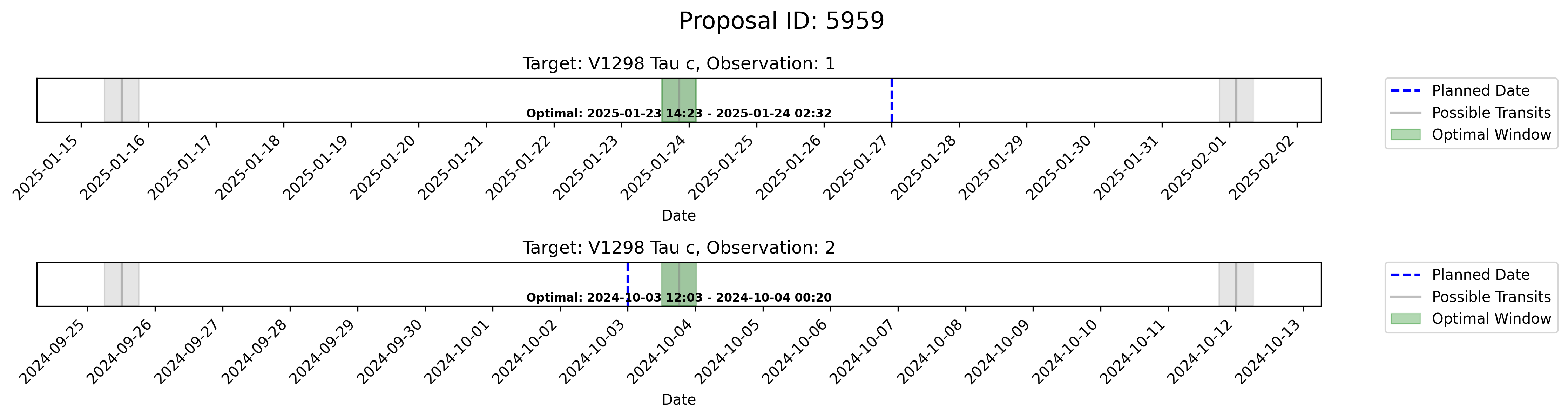}
    \caption{Timeline for Proposal 5959}
\end{figure}


\expandafter\input\expandafter{\jobname.bbl}


\begin{thebibliography}{99}

\bibitem{claimpin_overview_2010.7731E..07C}
M.~Clampin and E.~P. Smith.
Overview of the James Webb Space Telescope Observatory.
In \emph{Space Telescopes and Instrumentation 2010}, volume 7731, page 773107, 2010.

\bibitem{doyon_near_2023}
R.W. Doyon et al.
The Near Infrared Imager and Slitless Spectrograph for the James Webb Space Telescope.
\emph{PASP}, 135:098001, 2023.

\bibitem{feinstein_eleanor_2019}
A.D. Feinstein et al.
eleanor: An Open-source Tool for Extracting Light Curves from the TESS Full-frame Images.
2019.

\bibitem{gardner_james_2006}
J. Gardner et al.
The James Webb Space Telescope.
\emph{Space Sci Rev}, 123:485--606, 2006.

\bibitem{hippke_tls_2019}
M. Hippke and R. Heller.
TLS: Transit Least Squares.
Astrophysics Source Code Library, ascl:1910.007, 2019.

\bibitem{hord_uniform_2021}
B. Hord et al.
A Uniform Search for Nearby Planetary Companions to Hot Jupiters in TESS Data.
\emph{AJ}, 162:263, 2021.

\bibitem{Lightkurve2018}
Lightkurve Collaboration et al.
Lightkurve: Kepler and TESS time series analysis in python.
Astrophysics Source Code Library, 2018.

\bibitem{mcelwain_james_2023}
M. McElwain et al.
The James Webb Space Telescope mission: optical telescope element design, development, and performance.
\emph{PASP}, 135:058001, 2023.

\bibitem{ricker_transiting_2015}
G.R. Ricker et al.
Transiting Exoplanet Survey Satellite (TESS).
\emph{J. Astron. Telesc. Instrum. Syst.}, 1:014003, 2015.

\bibitem{rieke_performance_2023}
M.J. Rieke et al.
Performance of NIRCam on JWST in Flight.
\emph{PASP}, 135:028001, 2023.

\bibitem{Rigby_2023}
J. Rigby et al.
Title not provided.
2023.

\bibitem{stevenson_2016}
K. B. Stevenson et al.
Transiting Exoplanet Studies and Community Targets for JWST's Early Release Science Program.
\emph{PASP}, 128:094401, 2016.

\bibitem{sullivan_trans_2015}
P. W. Sullivan et al.
The Transiting Exoplanet Survey Satellite: Simulations of Planet Detections and Astrophysical False Positives.
2015.

\bibitem{Wright_2023}
G.S. Wright et al.
The Mid-infrared Instrument for JWST and Its In-flight Performance.
\emph{Publications of the Astronomical Society of the Pacific}, 135(1046):048003, May 2023.

\end{thebibliography}
\end{document}